# ALMA Memo 612

# ALMA BOARD

| ALMA EDM Document | AEDM 2018-017-O |
| --- | --- |
| Distribution | **Ordinary Session** |

**Subject**: The ALMA Development Roadmap

**Authors**: J. Carpenter, D. Iono, L. Testi, N. Whyborn, A. Wootten, N. Evans
(The ALMA Development Working Group)

**Purpose of Document**: To provide a public version of the ALMA Development Roadmap

**Status**: Approved by the Board by written procedure pursuant Art. 11 of the Board's Rules of Procedure

## Preamble

ALMA is approaching completion of its originally envisaged capabilities and, within the first five years of operations, the original fundamental science goals of ALMA have been essentially achieved. The ALMA Board established a Working Group to develop a strategic vision and prioritize new capabilities for the Observatory out to 2030 as part of the ALMA Development Program. The ALMA Board approved the resulting ALMA Development Roadmap in November 2017. The following document is a summary of the Roadmap, approved by the Board in June 2018 for broad distribution.

According to the vision in the Board-approved Roadmap, the current development priorities as based on scientific merit and technical feasibility, are:

- To broaden the receiver IF bandwidth by at least a factor two, and
- To upgrade the associated electronics and correlator.

These developments will advance a wide range of scientific studies by significantly reducing the time required for blind redshift surveys, spectral scans, and deep continuum surveys. In order of scientific priority, receiver upgrades are recommended for intermediate (200-425 GHz), low (< 200 GHz), and high (> 425 GHz) frequencies.

The Board acknowledges that there are other potential development areas for the future for which the science cases and technical feasibility need to be further demonstrated.

Sean M. Dougherty
ALMA Director

Toshikazu Onishi
ALMA Board Chair

June 28th, 2018

# Executive Summary

The present document outlines a roadmap for future developments that will significantly expand ALMA's capabilities and enable it to produce even more exciting science in the coming decades. The proposed developments are motivated by the groundbreaking results achieved by ALMA during its first five years of operation. We offer a vision of new and expanded science drivers surpassing the original fundamental science goals of ALMA, which have been essentially achieved in the first five years of ALMA operations. The roadmap described here is based on input on new scientific directions and technical feasibility of future developments from the ALMA Science Advisory Committee (ASAC), the community, and technical documents.

The Working Group recommends that the top development priority, based on scientific merit and technical feasibility, is to broaden the receiver IF bandwidth by at least a factor two, and to upgrade the associated electronics and correlator. These developments will advance a wide range of scientific studies by significantly reducing the time required for blind redshift surveys, chemical spectral scans, and deep continuum surveys. In order of scientific priority, receiver upgrades are recommended for intermediate (200-425 GHz), low (< 200 GHz), and high (> 425 GHz) frequencies.

The Working Group recommends that the receiver and throughput developments proceed as soon as fiscally and technically feasible. As a first step, a technical and scientific group should be formed to formalize the top-level requirements. A team of systems engineers should then be charged with flowing these requirements down to the sub-systems to form a consistent new set of minimum requirements, which future development projects would have to meet. Given that upgrading the throughput will impact many ALMA subsystems, the Working Group recommends that a team within ALMA be charged with coordinating and monitoring these developments.

The Working Group recommends that a group be tasked with prioritizing the long-term capabilities of the ALMA archive. The main emphasis should be to identify the functionality needed for the community to mine the ALMA archive efficiently, especially in view of the increased receiver bandwidth envisioned by the recommended developments.

The Working Group supports continued exploration of the following development paths which have potentially large impacts on ALMA science, but for which the science case and technical feasibility require further investigation.

- Extending the maximum baseline length by a factor of 2-3 provides the exciting opportunity to image the terrestrial planet forming zone in nearby protoplanetary disks. We recommend that studies commence to quantify the science drivers, and to investigate the technical requirements and logistical issues in implementing longer baselines.
- Focal plane arrays could significantly increase ALMA's wide-field mapping speed. We recommend that studies be supported to identify the compelling science cases, which can be used to define the technical requirements (e.g., preferred receiver bands, number of pixels, and bandwidth).
- Increasing the number of 12-m antennas would benefit all science programs by improving the sensitivity and image fidelity. We recommend that studies be supported that develop specific science cases for adding more antennas.
- A large single dish submillimeter telescope of diameter of at least 25-m would enable deep, multi-wavelength images of the sky and provide many scientific synergies with ALMA. However, we believe that the mode of operation of a large single dish telescope is not within the scope of current ALMA operations, and that a re-definition of the scope of the ALMA project by the ALMA Partners would be necessary to consider a large single dish telescope as appropriate for the ALMA development line.

# Table of Contents

# 1 Introduction

The ALMA Development Program supports the continued development and upgrades of hardware, software, and analysis tools for the ALMA project. Until now these funds have been used mainly to complete the complement of receiver bands envisioned in the ALMA Construction Project Book [1]. As this process moves toward completion, it is timely to define a new long-term development strategy for the upcoming decade and beyond, which will guide further advancements in ALMA's technical and scientific capabilities.

As a first step, the ALMA Science Advisory Committee (ASAC) examined potential technical developments for ALMA between now and the year 2030. As summarized in a collection of studies referred to as the "ALMA2030" report, that was submitted to the ALMA Board in March 2015, the ASAC recommended, with no specific priority, four development paths based on their long-term scientific potential [2]:

1. Improvements to the ALMA Archive: enabling gains in usability and impact for the observatory;
2. Larger bandwidths and better receiver sensitivity: enabling gains in speed;
3. Longer baselines: enabling qualitatively new science;
4. Increasing wide field mapping speed: enabling efficient mapping.

In order to review and prioritize the recommendations from the ALMA2030 report, the ALMA Director was charged by the ALMA Board to form a Development Working Group (hereafter Working Group) in November 2015. The Working Group originally consisted of the ALMA Director as chair, the JAO Systems Engineer, the JAO Observatory Scientist, a representative from the ALMA Board and the three regional Program Scientists. In April 2017, the ALMA Board appointed an acting Director, who delegated the leadership of the Working Group to the JAO Observatory Scientist.

The Working Group tasked itself with addressing two charges:

- On behalf of the community, the Working Group shall propose a science-driven vision for the medium (5 years) to longer-term (5 to 15 years) developments of ALMA.
- This resulting plan should be prioritized, remaining commensurable within the anticipated ALMA development budget.

In March 2017, the ALMA Director's Council encouraged the Working Group to also consider major advances, motivated by science, even if they required resources beyond the current nominal development budget.

The present document reports the findings and recommendations from the Working Group for future ALMA developments. Besides scientific potential, the recommendations incorporate technical readiness and feasibility, operational impact, organizational challenges and anticipated cost. The document is organized as follows. Section 2 describes the process used to review, select and prioritize the recommendations and developments discussed in the ALMA2030 report and to consider additional potential developments. Section 3 presents the new science goals that sustain the development vision. Section 4 summarizes the development path for upgrading the ALMA receivers, correlator and digital system, which is the top development priority recommended by the Working Group. Section 5 summarizes the Working Group's recommendation for further archive developments. Section 6 describes other developments that could potentially have long-term scientific impact for the observatory, but that need further study to develop the science case and determine their technical feasibility.

# 2 Process

The Working Group received input on the scientific directions and technical feasibility of future developments from the ASAC, the community, and technical reports.

The ALMA Science Advisory Committee (ASAC) summarized the scientific themes and overall development directions over the upcoming decade in the ALMA2030 report, which consists of four documents: *ASAC Recommendations for ALMA 2030* [2], *Pathways to Developing ALMA* [3], *Major Science Themes in the 2020-2030 Decade* [4], and *Major Facilities by 2030* [5]. Subsequently, the ASAC provided additional scientific guidance by prioritizing which receiver bands should be upgraded based on expected scientific impact. The ASAC also provided further input on the scientific case of expanding to longer baselines and new top-level science goals.

The Working Group received advice on the technical feasibility and readiness of various developments from the regional communities and development teams. The ALMA development program, in particular, has funded studies and projects that have examined possible upgrades to receivers, correlators, extended baselines, and the archive that address the recommendations in the ALMA2030 report. Table 1 lists the pertinent studies that have been completed or are in progress.

In addition, each of the ALMA Executives hosted a regional development workshop in the summer of 2016, which was attended by one or more members of the Working Group. The workshops provided an overview of the ongoing Studies and Projects in each region and the prospects of long-term developments. The list of workshops and links to the online presentations are provided in Table 2.

The Working Group presented a summary of the near, mid, and long-term developments in a special session at the *Half a Decade of ALMA: Cosmic Dawns Transformed* conference in Indian Wells, California in September 2016. The session included a 30 minute question-and-answer period to receive community feedback on the future ALMA developments.

The Working Group has expanded upon the ALMA2030 recommendations to also consider the addition of 12-m antennas within the current concession and the construction of a large single-dish submillimeter telescope as potential development directions for ALMA. These developments, while discussed in the ALMA2030 report, were not listed in the final ALMA2030 recommendations [2] and not discussed at the Indian Wells meeting.

**Table 1: ALMA Studies Pertinent to the ALMA2030 Recommendations**

| Authors | Date | Report |
|---|---|---|
| **Receivers** | | |
| Henke et al. [6] | March 2015 | Concept Study of a Millimeter Camera for ALMA |
| Withington et al. | April 2015 | Supra-THz interferometry with ALMA |
| Kerr et al. [7] | March 2016 | Towards a Second Generation SIS Receiver for ALMA Band 6 |
| Kerr et al. [8] | March 2016 | Towards a Second Generation SIS Receiver for ALMA Band 10 |
| Hesper et al. | September 2016 | Overview of current ALMA Band 9 2SB mixer results |
| Cyberey & Lichtenberger [9] | March 2017 | Advanced Materials and On-wafer Chip Evaluation: 2nd Generation |
| Kojima et al. [10] | October 2017 | Performance and Characterization of a Wide IF SIS-Mixer-Preamplifier Module Employing High-Jc SIS Junctions |
| Henke, Niranjanan, & Knee | In progress | Prototype of a Complete Dual-Linear 2SB Block and a Single-Polarization Balanced 2SB Block |
| Kerr et al. | In progress | Development of 2nd Generation SIS Receivers for ALMA |
| Mroczkowski & Yagoubov | In progress | The Case for a Combined Band 2+3 Receiver |
| **Extended baselines** | | |
| Kameno et al. [11] | October 2013 | ALMA Extended Array |
| **Digitizers** | | |
| Quertier et al. | December 2017 | Digitizer upgrade |
| **Correlator** | | |
| Escoffier et al. [12] | March 2015 | Enhancing the Spectral Performance of the 64-antenna ALMA Correlator |
| Escoffier et al. [13] | May 2015 | Doubling the Bandwidth of the 64-Antenna ALMA Correlator |
| Lacasse et al. [14] | January 2017 | Spectral Resolution and Bandwidth Upgrade of the ALMA Correlator |
| Baudry et al. [15] | July 2017 | Digital Correlation and Phased Array Architectures for Upgrading ALMA |

**Table 2: Working Group Participation in ALMA Workshops and Conferences**

| Workshop / Conference | Dates | Webpage |
|---|---|---|
| European Development Workshop | May 25-27, 2016 | Link |
| East Asia Development Workshop | July 20-21, 2016 | Link |
| NRAO Development Workshop | August 23-25, 2016 | Link |
| Half a Decade of ALMA: Cosmic Dawn Transformed | September 20-23, 2016 | Link |

## 3 New Fundamental Science Drivers

The three level-one science goals of the ALMA Baseline project have been essentially achieved in the first five years of ALMA operations.

- *The ability to detect spectral line emission from CO or $C^+$ in a normal galaxy like the Milky Way at a redshift of z = 3, in less than 24 hours of observation;*
- *The ability to image the gas kinematics in a solar-mass protoplanetary disk at a distance of 150 pc, enabling one to study the physical, chemical, and magnetic field structure of the disk and to detect the tidal gaps created by planets undergoing formation;*
- *The ability to provide precise images at an angular resolution of 0.1"*

The Working Group proposes the following fundamental science drivers for ALMA developments over the next decade:

**Origins of Galaxies**

Trace the cosmic evolution of key elements from the first galaxies (z>10) through the peak of star formation (z=2–4) by detecting their cooling lines, both atomic ([CII], [OIII]) and molecular (CO), and dust continuum, at a rate of 1-2 galaxies per hour.

**Origins of Chemical Complexity**

Trace the evolution from simple to complex organic molecules through the process of star and planet formation down to solar system scales (~10-100 au) by performing full-band frequency scans at a rate of 2-4 protostars per day.

**Origins of Planets**

Image protoplanetary disks in nearby (150 pc) star formation regions to resolve the Earth forming zone (~ 1 au) in the dust continuum at wavelengths shorter than 1mm, enabling detection of the tidal gaps and inner holes created by planets undergoing formation.

Achieving these ambitious goals is currently impossible even with the outstanding capabilities of the current ALMA array. These science goals can be achieved with the upgrades proposed in this document, upgrades that would make ALMA even more powerful and keep it at the forefront of astronomy by continuing to produce transformational science and enabling fundamental advances in our understanding of the universe for the decades to come.

## 4 Upgrading the Receivers, Digital System, and Correlator

The Working Group recommends that the top development priority for ALMA should be to expand the bandwidth of the receivers and upgrade the digital system and correlator. The current ALMA digital processing signal and correlator handle 16 GHz of bandwidth[1] (8 GHz per polarization). However, the 8 GHz instantaneous frequency coverage per polarization covers a small fraction of the atmospheric windows. Therefore, with the current system, blind redshift programs and astrochemical surveys need multiple tunings to cover an appreciable range of redshifts or to obtain an extensive chemical inventory. Expanding the throughput by a factor of two or more is required to achieve the new science goals (see Section 3) by reducing the time to conduct blind redshift surveys or chemical spectral scans by at least a factor of two and a factor of 8-16 for high spectral-resolution spectral scans, increasing the ability to observe critical molecular transitions simultaneously, and improving the continuum sensitivity to image disks at high spatial resolution.

[1] Bands 9 and 10 are DSB receivers, and the lower and upper sideband signals can be separated using 90º phase switching.

Significant gains in the observing speed can be achieved by upgrading i) the receivers to deliver larger IF bandwidths, ii) the digitizers and digital processing to allow for larger basebands and higher effective bandwidth coverage, and iii) the correlator to process larger bandwidths at higher spectral resolution. The potential gains are substantial:

- The speed in which a spectral region can be imaged is proportional to the bandwidth (*B*) and system temperature ($T_{sys}$) as $B/T_{sys}^2$. Thus doubling the bandwidth doubles the imaging speed for spectral line surveys and continuum observations assuming no loss in receiver sensitivity.
- The analog IF signal is split into 2 GHz basebands, with effective bandwidth coverage within these frequency ranges of approximately 94%. Reducing or eliminating the coverage gap will improve the completeness or efficiency of broadband spectral scans.
- Narrow-line sources require high spectral resolution, which currently can only be achieved by reducing the bandwidth processed by the correlator by a factor of two or more. For example, Band 7 spectral scans at a spectral resolution of 0.25 km/s, a typical resolution desired for disk and star formation science, requires ~20-30 individual setups using 0.47 GHz wide windows, which is a factor of 4 smaller than the full baseband. Thus processing the full IF bandwidth at high spectral resolution can increase the efficiency of spectral scans by factors of several.

The current performance and potential upgrades of the ALMA receivers is discussed in Section 4.1. The recommended prioritization for the receiver upgrades is provided in Section 4.2.

## 4.1 Overview of Current ALMA Receivers and Potential Upgrades

Table 3 lists the IF bandwidth and receiver type for the current suite of receivers on the ALMA antennas, as well as the ongoing (Bands 1 and 5) and anticipated (Band 2) receiver developments. The intermediate frequency bands (Bands 3 to 8) have 4-5 GHz of IF bandwidth with sideband separating mixers. The higher frequency bands (Bands 9 and 10) are double sideband systems, in which both sidebands will be recorded simultaneously, starting in Cycle 5 with the implementation of 90° phase switching.

Figure 1 shows the receiver temperature for the current ALMA receivers. At low frequencies (Bands 3 to 7), the receiver noise temperatures are just a few times the quantum limit and approach practical physical limits. This suggests the most significant gains in performance at the lower frequency bands will mainly involve increasing the IF bandwidth, while the higher frequency bands can be improved in receiver temperature and/or wider IF bandwidth.

When designing wide IF receivers, the scientific gains obtained by expanding the IF need to be balanced against the potential increase of the receiver noise [16]. This was one of the original reasons for choosing the original 4 GHz IF bandwidth of the ALMA receivers. With the improvement of receiver technology, 8 GHz IF bandwidth receivers are now a competitive alternative and implemented at other observatories. For instance, NOEMA receivers deliver 31 GHz of bandwidth (7.7 GHz of bandwidth per polarization per sideband) in the equivalent of ALMA Bands 3, 4, and 6 [17]. The SMA also offers receivers with 8 GHz of bandwidth per sideband, with the possibility of providing 32 GHz of continuous frequency coverage using a pair of receivers simultaneously. Current technology may allow even broader bandwidths that could potentially cover the ALMA RF frequency ranges with fewer receiver bands and much increased IF bandwidth [10]. The SMA also plans to extend the IF to 16 GHz of bandwidth per sideband [18].

Any receiver upgrade that increases the IF bandwidth must be accompanied by an upgraded signal transport chain to digitize the receiver output and transfer the data to the correlator, an upgraded correlator to process the increased bandwidth, and an upgraded software and data flow system to process the data. Under the ALMA Development Program, ESO is funding an ongoing study to design and prototype new digitization and tunable filter processing electronics that would allow ALMA to present a broader IF band at the input of the correlator [19]. The system is currently being designed for 32 GHz bandwidth, split in 4 GHz basebands and with higher effective bandwidth coverage. The possibility of using higher bits digitization than the current ALMA system is being explored to improve the sensitivity, and also to increase the resilience to IF power variation of the receivers. Larger RF bandwidths may also allow two bands to be covered with a single receiver [20].

**Table 3: Current ALMA Receiver Characteristics**

| Band | IF (GHz) | Type |
|---|---|---|
| 1 (under construction) | 4-12 | SSB |
| 2 (under development) | 4-12 | 2SB |
| 3 | 4-8 | 2SB |
| 4 | 4-8 | 2SB |
| 5 (Cycle 5) | 4-8 | 2SB |
| 6 | 5-10 | 2SB |
| 7 | 4-8 | 2SB |
| 8 | 4-8 | 2SB |
| 9 | 4-12 | DSB |
| 10 | 4-12 | DSB |

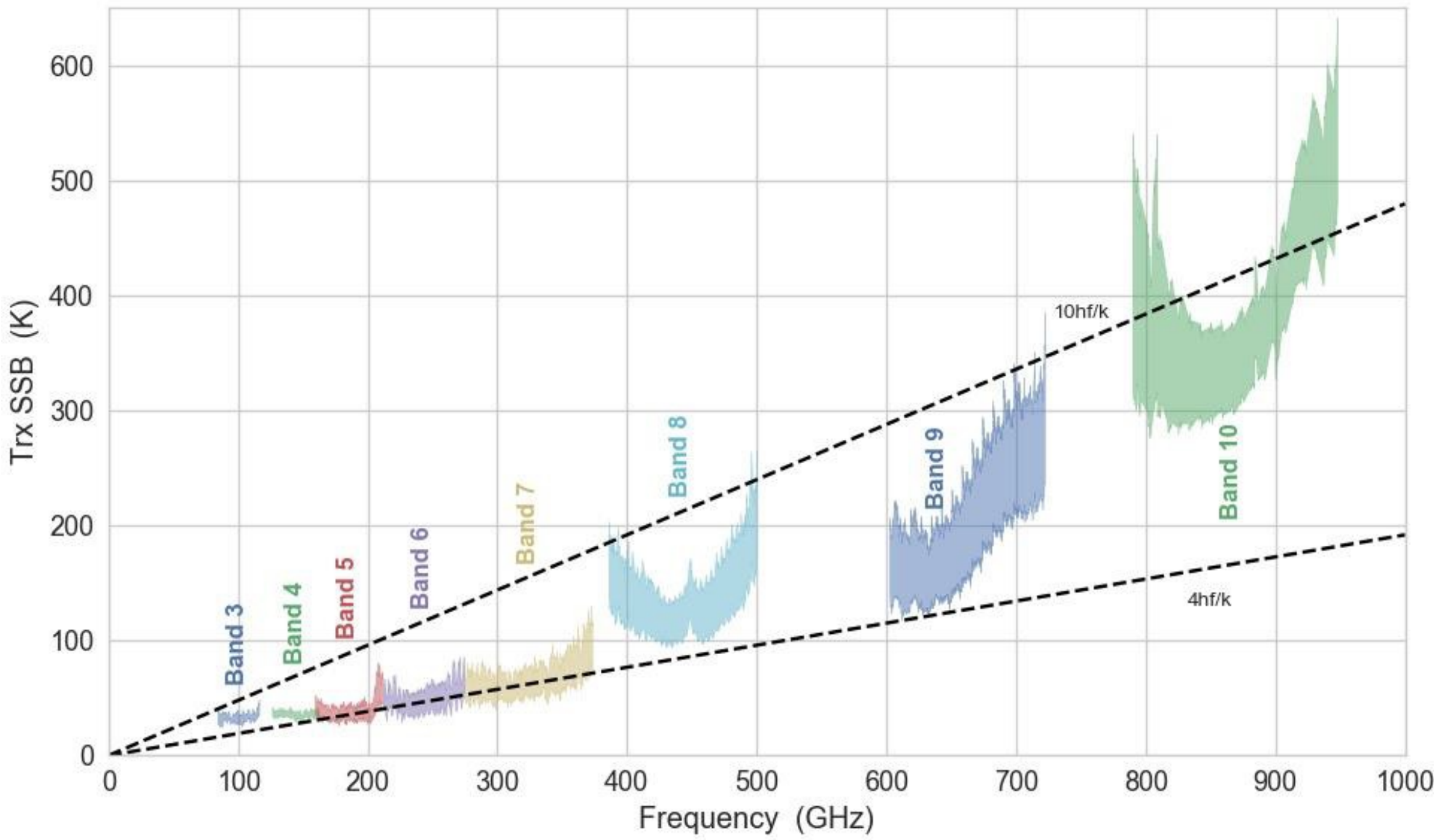

**Figure 1:** Receiver noise temperature for the ALMA receiver, where the shaded region encompasses 75% of the receivers about the median receiver temperature. Bands 3-8 are 2SB receivers, and Bands 9 and 10 are DSB. The noise temperature shown for the DSB receivers are twice the DSB temperatures. Band 1, which is under construction, has a receiver specification of < 25 K across 80% of the band.

## 4.2 Prioritization of Receiver Upgrades

The frequency ranges recommended for receiver upgrades, with the highest priority listed first, are intermediate (200-425 GHz), low (< 200 GHz), and high (> 425 GHz). These frequency ranges were prioritized because they have the most direct impact on the new science goals (see Section 3) as described below. The prioritization of receiver upgrades is based on scientific and technical analysis in 2017. Priorities may be revisited as either the science or technology evolves. In particular, innovative, cost effective technological developments should always be welcome.

*Origins of Galaxies*

- Intermediate and high frequencies are optimal for spectral surveys of fine structure lines to probe galaxies in the epoch of reionization and early galaxy evolution. Intermediate frequencies can probe [CII] (158 μm) at $z$=3.5-8.5, [NII] (205 μm) at z=2.4-6.3, and [NII] (121 μm) at $z$=4.8-11.3. High frequencies trace [OIII] (52 μm) at $z$=5.1-12.6, [OIII] (88 μm) at z=2.6-7.0, and [OI] (63 μm) at $z$=4.0-10.2.
- High frequencies probe [CII] from $z$=1.0 to 3.5, bracketing the peak of the star formation rate in the Universe.
- With an 8 GHz IF receiver, the low frequencies can be covered much more efficiently, which would significantly increase the efficiency of spectral surveys of high redshift galaxies to obtain a spectroscopic redshift and characterize the physical and chemical properties of the gas. These frequency ranges could probe galaxies at nearly all redshifts up to $z$=10 using a variety of molecular and atomic tracers.

*Origins of Chemical Complexity*

- Intermediate frequencies are essential for most astrochemical studies because they contain the brightest lines of both dense gas tracers and complex organic molecules, considering both intrinsic line strengths and excitation conditions. Low frequencies are needed for cold galactic sources of astrobiology interest (e.g., pre-stellar cores). A doubling of bandwidth, together with an increase in spectral channels, would dramatically increase the efficiency of spectral surveys toward star forming regions and protoplanetary disks, by reducing the required observing time by a factor of 8, assuming ~0.25 km/s resolution can be achieved across the entire bandwidth.
- With increased bandwidth, several specific rotational transitions of CO, $^{13}CO$ and $C^{18}O$ could be covered simultaneously, which is key to constraining gas masses from protoplanetary disks to galaxies.

*Origins of Planets*

- Intermediate frequencies are the choice for most continuum studies, because of a combination of good phase stability (giving access to long baselines), high intrinsic surface brightness and high spatial resolution compared to lower frequency bands. Doubling the bandwidth would increase the observing efficiency and expand the sample of sources that can be self-calibrated and therefore observed with long baselines.
- High frequencies provide access to the smallest spatial scales when used with the 16 km baselines. An upgrade of this band would optimize observations of protoplanetary disks at 1 au spatial resolution in the nearest star forming regions, including perhaps accreting protoplanets. Upgrading the high frequency receivers will also enable self-calibration for fainter sources at longer baselines.
- Combined with extended baselines (see Section 6.1), the intermediate frequencies will provide approximately 1 au spatial resolution in nearby circumstellar disks.

# 5 Archive Development

The Working Group concurs with the ALMA2030 report that the ALMA Science Archive will become the primary source for an increasing number of publications. The ability to efficiently mine the archive contents is therefore vital for the community and ALMA's future. With the expansion of the receiver bandwidth and the upgraded correlator envisioned here, not only will the archive capacity need to be increased, but also the capabilities will need to be enhanced to exploit the rich repository of spectra.

The current ALMA archive provides the basic functionality to search and download data, and the complementary Japanese Virtual Observatory (JVO) provides impressive functionality to visualize data cubes. In addition, the ALMA Archive Working Group has produced a 5-year development plan that will provide additional functionality to facilitate archival research. While the cost of further archive upgrades is likely much less than envisioned for the package of receiver, digital system, and correlator, careful planning is nonetheless required to anticipate the needs of the community.

The Working Group recommends that a committee be formed and tasked with prioritizing the scientific functionality that will be needed in the ALMA archive over the next decade, with particular attention toward mining the archival spectra produced by the receiver upgrades.

# 6 Mid-term Opportunities

The ALMA2030 report recommended expanding the maximum baselines to improve the angular resolution and developing focal plane arrays to improve the wide-field mapping speed. The Working Group further considered the scientific impact of adding additional antennas to improve the sensitivity and *u,v* coverage, and of a large, single-dish submillimeter telescope that can rapidly survey the sky. Each of these developments is explored in this section. In general, the Working Group recommends that before ALMA can commit to any of these developments, the science case and technical feasibility need more careful consideration. If mature, these initiatives may be ready for further considerations by the middle of the 2020s.

## 6.1 Extended Baselines

A top-level science goal for ALMA is to study the physical, chemical, and magnetic field structure of protoplanetary disks and to detect the tidal gaps created by planets undergoing formation. The spectacular images of HL Tau, TW Hydra and Elias 27 among others have arguably surpassed even the optimistic expectations, even though ALMA has yet to reach full capabilities.

The ALMA image of dust continuum in TW Hydra was particularly noteworthy since it reaches a spatial resolution of 1 au to resolve the terrestrial planet zone. However, as TW Hydra is the closest circumstellar disk, approximately 3 times better angular resolution (~7 mas) is needed to obtain 1 au spatial resolution in the nearest star-forming regions; i.e., Ophiuchus, Lupus, and Taurus. This can be obtained by either observing in high frequencies with the existing 16 km baselines by developing an advanced method of atmospheric phase correction, or in intermediate frequencies with 2 to 3 times longer baselines. Extending the baselines by two to three times (30 to 50 km) is scientifically advantageous given the lower dust optical depths at lower frequencies within the disk. In addition, the lower frequency bands can be observed more frequently.

Extending the ALMA baselines by a factor of 2-3 would require the construction of new antenna pads and associated infrastructure, as well as new antennas, located well outside of the current ALMA concession. Transporting antennas from the existing array to and from the remote pads of the extended baseline array will not be feasible due to the difficulties associated with transporting the antennas over such long distances, the need to cross or use public highways, and the substantial cost of constructing roads to the standard necessary for use by the transporter. The lengths of the baselines therefore require antennas permanently stationed on the distant pads.

The Working Group assumed that at least six additional antennas, permanently stationed on new pads, would be needed to provide minimal *u,v* coverage on the longest baselines. Detailed studies are needed to determine if this number is sufficient to achieve the necessary image fidelity and sensitivity in brightness temperature to probe 1 au scales in nearby protoplanetary disks. Detailed topographical studies are also needed to identify suitable locations for the antenna pads. Further studies are required to estimate how frequently the very long baselines will be used and decide what role the remote antennas will play when the array is in compact configurations. Finally, intermediate configurations, with baselines up to 20-25 km, should be studied as a step toward 30-50 km baselines. Such an intermediate configuration may also be simpler in terms of logistics.

The Working Group recommends that the development program support studies of the scientific, technical, and logistical issues involved in an increase in maximum baseline length.

## 6.2 Focal Plane Arrays

Focal plane arrays could significantly increase ALMA's wide-field mapping speed to survey large regions of molecular clouds, image nearby galaxies, and conduct deep-field cosmological surveys. Existing heterodyne multi-beam arrays typically provide 4-16 elements with on-sky footprints of 1−6 arcmin$^2$, with the largest arrays containing 64 elements. Such receivers are likely to occupy a significant fraction of the (already tightly packed) available focal plane space in the ALMA antennas and are likely feasible for only one band at a time. Modest pixel counts (perhaps 4 to 16) can likely be accommodated on ALMA without major redesign of the antenna optics. A receiver array covering four times the area of a single beam for an extended source cuts integration times by factor of four, assuming that individual receiver detectors do not suffer any performance loss within their array configuration, and that the required bandwidth is not compromised relative to a single pixel. In particular, the latter requirement would need for the ALMA correlator to have an excess bandwidth capacity compared to a single pixel receiver.

The Working Group finds that both the science case and the technical feasibility of focal plane arrays needs to be significantly more advanced before ALMA can consider their implementation. In particular, the science case needs to define the preferred band, the number of pixels, and bandwidth that will motivate technical studies. The technical and scientific tradeoffs involved in developing and using multi-pixel receivers in ALMA are complex and require further investigation to evaluate feasibility, and will certainly have a significant impact on observatory operations. Upgrading even one band to a multi-pixel receiver requires a number of improvements in elements downstream (IF transport, correlator, archive), and probably upstream (LO distribution).

The Working Group recommends that the development program support studies of the scientific, technical, and logistical issues involved in implementing focal plane arrays.

## 6.3 Additional 12-m Antennas

Adding 12-m antennas to the baseline array would benefit all science programs by increasing the sensitivity and/or decreasing the integration time, while improving the image fidelity and quality. The longest baseline configurations would have significantly improved *u,v* coverage to image the complicated emission regions with high resolution, especially high frequencies. Therefore, increasing the number of 12-m antennas would help to achieve the new ALMA science drivers while making use of the full potential of the ALMA site.

The original ALMA construction plan called for 64 12-m diameter antennas, which was later de-scoped to 50 antennas for the 12-m array as a cost-savings measure. The correlator requirements were retained, and thus an additional 14 antennas could be added without significantly impacting the rest of the system.

Adding additional antennas also brings operational and imaging capabilities that have significant scientific benefits. With more antennas, the antenna configurations, originally designed for 64 antennas, can cover a wider range of resolution so that a greater number of programs can be accommodated with a given configuration, saving time needed now for reconfigurations and utilizing more easily the best weather conditions during the year. The better sensitivity can make more calibrators accessible closer to the source and allow for better phase correction, which is especially important for long baseline and high frequency observations. The robustness and availability of the self-calibration technique will be increased since the number of baselines will increase even more ($\sim N^2$) than the number of antennas ($\sim N$).

The Working Group recommends that the development program support studies that develop specific science cases for an increased number of antennas.

### 6.4 Large Single-dish Submillimeter Telescope

A large, single-dish submillimeter telescope of diameter between 25 to 50 meters would provide powerful scientific synergies with ALMA. Such a telescope will survey the sky in the submillimeter continuum thousands of times faster than ALMA to identify large samples of galactic and extragalactic sources.

Multi-wavelength submillimeter cameras allow for effective extragalactic spectral surveys over wide areas of the sky. Such surveys provide a 3D mapping of the deep Universe in the submillimeter at moderate angular resolution. Highly multiplexed panoramic arrays for heterodyne spectroscopy are already being developed and a large single dish could efficiently map the Milky Way and nearby galaxies. Another key science application for a large single-dish submillimeter telescope would be the detection and characterization of large-scale polarization of the interstellar medium at moderate resolution. These and other science cases have been reviewed in the ESO Submillimeter Single Dish Scientific Strategy Working Group Report. In all of these cases, ALMA would be perfectly suited to follow up individual sources to detect fainter lines or to resolve source structure.

While noting the powerful synergies with ALMA, the Working Group believes that the mode of operation of a large single-dish telescope, with presumed emphasis on large scale sky surveys that are independent of direct interferometer observations, is not within the scope of current ALMA operations. A re-definition of the scope of the ALMA project by the ALMA Partners would be necessary to consider a large single-dish telescope as appropriate for the ALMA development line.

## 7 Concluding Remarks

The main recommendation of the Working Group is that ALMA initiates as soon as possible a program to broaden the IF bandwidth and increase the overall data throughput. These developments will have profound impacts on many aspects of the ALMA activities that will require a sustained program over the next decade. Careful planning is mandatory.

The Working Group recommends that a technical and scientific group be formed to formalize the top-level requirements. A team of systems engineers should then be charged with flowing these revised requirements down to the sub-systems to form a consistent new set of minimum requirements which future development projects would have to meet. Because of the long development timescale, the state-of-the-art in receiver bandwidth will likely evolve from the current 8 GHz IF. The top-level requirements for the digital system should be set in anticipation of future growth. This approach will require the regional development programs to reach consensus on the technical requirements. Further, because upgrading the throughput will require coordination of all the regions and the JAO, the Working Group strongly recommends that a team be set up within ALMA to oversee upgrade of the receivers, digital system, and correlator.